\documentclass[aps,nofootinbib]{revtex4}
\usepackage{epsfig}
\input epsf.tex
\usepackage[T1]{fontenc}
\usepackage[polish,english]{babel}
\usepackage{color}

\usepackage{graphicx}
\usepackage{amsmath}

\begin{document}
\title{Electromagnetically Induced Transparency and slow light
in media with Rydberg Excitons }
\author{Sylwia Zieli\'{n}ska-Raczy\'{n}ska$^1$,
 David Ziemkiewicz$^1$, Gerard Czajkowski$^1$}
\affiliation{$^1$ Institute of Mathematics and Physics, UTP
University of Science and Technology, Al. Prof. S. Kaliskiego 7,
PL 85-789 Bydgoszcz
 (Poland)}
\begin{abstract} We show that the Electromagnetically Induced Transparency (EIT) is
possible in a medium exhibiting Rydberg excitons and indicate the realistic parameters to
perform the experiment.  The
calculations for a Cu$_2$O crystal are given which show that in this medium due to large group index
one could expect slowing down a light pulse by a factor about $10^4$.\end{abstract}
\maketitle
\section{Introduction}

Recently, a lot of  attention has been directed to the subject of excitons in bulk crystals due to experimental observation of the so-called yellow exciton series
 in Cu$_2$O up to a large principal quantum number of $n = 25$ \cite{Kazimierczuk}-\cite{Zielinska.PRB.2016.b}. Such excitons in copper oxide,
  in analogy to atomic physics, have been named Rydberg excitons. By virtue of their special properties Rydberg excitons are of fascination
  in solid {state} and optical physics. These objects, whose size scales as the square of the Rydberg principal quantum number $n$, are ideally suited
   for fundamental quantum interrogations, as well as detailed classical analysis.
One could expect that Rydberg excitons would have been described,
in analogy to Rydberg atoms, by Rydberg series of hydrogen atoms,
but it turned out that this generic method of description should
has been revised because diameter of such exciton is much larger
then wavelength of light needed to create
it~\cite{Zielinska.PRB,Zielinska.PRB.2016.b}.

The observation and detailed description of Rydberg excitons have
opened a new field in condensed matter spectroscopy. In analogy to
medium of Rydberg atoms, where it has   been
possible to obtain a large optical nonlinearity at the single
photon level and to realize a lot of quantum optics sophisticated
experiments such as optical Kerr effect  or correlated states
\cite{Firstenberg}, it is expected that  the medium of Rydberg
excitons is also fertile area.
  The unique combination of their huge size, long radiative lifetimes, possible strong dipole-dipole interaction and,
what is the most important advantage for future technological applications, miniaturization of media/samples they are realized in, can be exploited to perform robust light-exciton quantum interfaces for quantum information processing purposes.

Solid bulk media are  systems  well worth considering for storing
quantum information because they have a number advantages over
gases, where a lot of experiments have been done [for recent
review see Ref.\cite{Firstenberg}]: they are easy to prepare,
diffusion processes are not so  fast, much higher densities of
interacting particles can be achieved  \cite{Johnsson}. A common
class of solids used within a quantum information context are
rare-earth-metal-doped crystals, where a long time over {one}
minute of storing information has been achieved  \cite{Heinze}, \cite{Schraft},
and nitrogen-vacancy centres in diamond \cite{nitrogen}
which have relatively long spin coherence. However, the size of
solid samples used in such experiments is of order of several
millimetres. Rydberg excitonic samples are much smaller:
observation of dipolar blockade in bulk Cu$_2$O
\cite{Kazimierczuk} and quantum coherence \cite{Gruenewald} were
performed in samples as small as several tens of micrometers.
Realization of these experiments have unlocked the plethora of
dynamical effect which might be observed in Rydberg excitons
media; one of examples is the electromagnetically induced
transparency, the performing of which in Cu$_2$O bulk crystal will
be the  next step toward potential implementation this medium for
quantum information processing.

Electromagnetically induced transparency (EIT)  \cite{Harris},
 is one of the important effect in
quantum optics as it allows for the coherent control of materials'
optical properties. The generic EIT bases on extraordinary
dispersive properties of an atomic medium with three active states
in the $\Lambda$ configuration. This phenomenon leads to the
significant reduction of absorption of a resonant probe, weak
laser field by irradiating the medium with a strong control field
making an otherwise opaque medium transparent. It leads to
dramatic changes of dispersion properties of the system:
absorption forms a dip called the transparency window and
approaches zero while dispersion in the vicinity of this region
becomes normal with a slope, which increases for a decreasing
control field. The resonant probe beam is now transmitted almost
without losses. EIT has been explained by destructive quantum
interference between different excitation pathways of the excited
state or alternatively  in terms of a dark superposition of
states. Since at least 20 years there has been a considered level
of activity devoted to EIT \cite{Fleischhauer}, which has been motivated by recognition
of a number of its applications among which slowing and storing
the light, see i.e., \cite{AR1}-\cite{AR3} and references therein,
 which allows for realization of delay lines and buffers
in optical circuits are well-known examples. A remarkable
quenching of absorption due to EIT in an undoped bulk of Cu$_2$O
in a $\Lambda$-type configuration involving lower levels was
examining in \cite{Artoni}. Demonstration of EIT in Rydberg atoms
involving the ladder levels scheme by Mohapatra \emph{et al}
\cite{Mohapatra}  has taken advantage of their unique properties.
In Rydberg systems the ladder configuration enables to couple
long-living metastable, initially empty upper level with the
levels coupled by the probe field. Rydberg
excitons in Cu$_2$O offer a great variety of accessible states for
creating the ladder configuration what enables such a choice of
coupling which could be realized by accessible lasers or eventually
to be suitable for desirable coherent interaction
implementations.
  Rydberg EIT
has also attracted attention with demonstration of
interaction-enhanced absorption imaging of Rydberg excitations
\cite{Gunter}. The non-linear response of Rydberg medium is
proportional to the group index and to the strength of the
dipole-dipole interaction and both of them can be extremely large.
EIT provides a possibility of dissipation-free sensing and probing
of highly excited Rydberg states and coherent coupling of Rydberg
states via EIT could be used for cross-phase modulation and photon
entanglement.

Rydberg excitons offer an unprecedented potential to study above mentioned
phenomena in solids. It seems that, similar to Rydberg atoms, strong dipolar interaction between
Rydberg excitons could appear and leads to so-called photon-blockade,
which offers promising means to tailor deterministic
single-photon sources \cite{Peyronel},
and to realize photonic phase gates \cite{Paredes}.
Up till now the researches involving in
Rydberg excitons, both theoretical and experimental,
have concentrated on their static properties
(excitonic states, electro- and magneto-optical properties).
The first step toward study the photon blockade, quantum non-linear optics,
 and many-body physics with Rydberg excitons is the realization of Rydberg EIT.
 Here we focuse our interest on  dynamical aspects and properties of
Rydberg excitons in Cu$_2$O and show, that the Rydberg excitonic states
can be used to perform the EIT. We indicate excitonic states which guarantee
 of the most efficient realization of the experiment taking into
account our previous results concerning excitonic resonaces as well as
damping parameters and the  matrix elements of momentum operator.

Our paper is organized as follows. In Sec. II we present the
assumptions of the considered model and solve the time evolution
equations, obtaining an analytical expression for the
susceptibility and the group index. We use the obtained expression
to compute the real and imaginary part of the susceptibility and
the group index (Sec. III), for a Cu$_2$O crystal slab. We examine
in details the changes of both real and imaginary part caused by
changes of driving parameters (for example, the Rabi frequencies).
In Sec. IV we draw conclusions of the model studied in this paper
and indicate the optimal choice of Rydberg states to realize the
EIT  and light slowing. The possible practical importance is also
indicated.
\section{Theory}
\begin{figure}
\includegraphics[width=.45\linewidth]{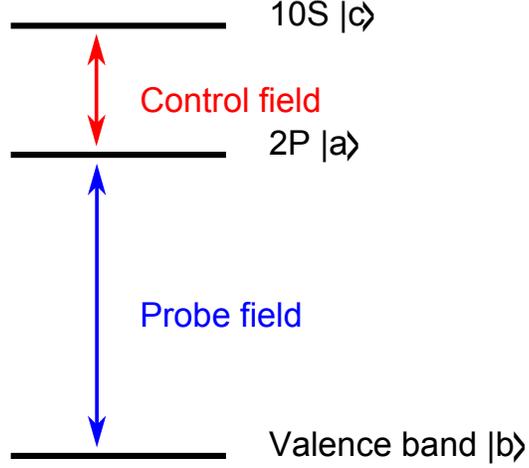}
\caption{Schematic of the considered ladder EIT system.}\label{Fig_schemat}
\end{figure}

The phenomenon of EIT, described qualitatively above, has been
studied theoretically for various configurations of the
transitions and probe and control beams. Below we propose a
theoretical description for the case when the atomic transitions
are replaced by intra excitonic transitions in Rydberg excitons
media. Condensed matter exhibits quite a variety of three-level
systems where induced transparency could be achieved in much the
same way as done with atoms. Yet dephasings, which can easily
break the coherence of the population trapping state, are
typically much faster in solids than in atomic vapors; it has
caused  difficulty to observe a large electromagnetic induced
transparency effect in solids. This  difficulty can be overcame by using
the Rydberg exciton states. Higher states have  much larger
lifetimes and the dephasing can reach the value which  enables
 observation of the EIT effect. Below we will consider a Cu$_2$O crystal as a
medium where the EIT phenomenon can be realized. We  use the ladder
 configuration (Fig. \ref{Fig_schemat}), consisting of three
levels $a, b$, and $c$. As in previous works on Rydberg excitons,
we focus our attention on the so-called yellow series associated
with the lowest inter-band transition between the $\Gamma^+_{7}$
valence band and the $\Gamma^+_{6}$ conduction band. Because both
band-edge states are of even parity, the lowest $1S$ exciton state
is dipole-forbidden, whereas all the \emph{P} states are
dipole-allowed; the $1S$ to $nP$ transition is also allowed. We
have chosen the valence band as the $b$ state. The $10S$ and $2P$
excitonic states are the $a$ and $c$ states, respectively. To
obtain the $S$ state we assume that a constant electric field $F$
is applied to the considered system. As it was recently shown, the
applied field splits the $P$ levels into $S$, $P$, and $D$
states\cite{Zielinska.PRB.2016.b}.

Let the probe/signal field of
frequency $\omega_1$ and amplitude $\varepsilon_1$ couples the
ground state $b$ of energy $E_b$ with an excited state $a$ of
energy $E_a$. The control field of frequency $\omega_2$  and
amplitude $\varepsilon_2$ couples a state $c$ of energy $E_c$ with
the state $a$, as it is illustrated in Fig. \ref{Fig_schemat}.  The Hamiltonian of
such three level system interacting with an electromagnetic wave,
in the rotating wave approximation reads
\begin{eqnarray}
H&=&E_{a}\vert a\rangle\langle a\vert+E_{b}\vert b\rangle\langle b\vert+E_{c}\vert c\rangle\langle c\vert\nonumber\\
&&+\left\{-\hbar\Omega_{1}(z,t)\exp[-{\rm i}(\omega_{1}t-k_{1}z)]\vert a\rangle\langle b\vert\right.\\
&&\left.-\hbar\Omega_{2}(z,t)\exp[-{\rm
i}(\omega_{2}t-k_{2}z)]\vert a\rangle\langle
c\vert+h.c.\right\},\nonumber
\end{eqnarray}\\
where $k_1$, $k_2$,
$\Omega_{1}(z,t)=({1}/{\hbar})d_{ab}\varepsilon_{1}(z,t)$,\,\,\,
$\Omega_{2}(z,t)=({1}/{\hbar})d_{ac}\varepsilon_{2}(z,t)$ are
the wave vectors and Rabi frequencies corresponding to the particular
couplings  and $d_{ij}$ being the dipole transition moments
related to the specific transitions. The time evolution of the system is governed
by the von Neumann equation with a phenomenological relaxation
contribution
\begin {equation}
{\rm i}\hbar\frac{{\rm d}\rho}{{\rm d}t}=[H,\rho]+R\rho,
\end {equation}
where $\rho(z,t)$ denotes the density matrix for an exciton at the
position $z$, and $R$~is the relaxation operator accounting for all
relaxation processes in the medium.
Setting
\begin{eqnarray}
\rho_{ab}&=&\sigma_{ab}\exp[-{\rm i}(\omega_1 t-k_1 z)],\nonumber\\
\rho_{ac}&=&\sigma_{ac}\exp[-{\rm i}(\omega_{2}t-k_2 z)],\nonumber\\
\rho_{bc}&=&\sigma_{bc}\exp[-{\rm i}(\omega_1-\omega_2)t-(k_1-k_2)z)],\nonumber\\
\rho_{aa}&=&\sigma_{aa},\nonumber\\
\rho_{bb}&=&\sigma_{bb},\nonumber\\
\rho_{cc}&=&\sigma_{cc},\nonumber
\end{eqnarray}
one can get rid of time-dependent factors except for slowly
varying factors accompanying probe field.

We denote by
$\delta_1=(E_a-E_b)/\hbar-\omega_{1}$ and
$\delta_2=(E_a-E_c)/\hbar-\omega_{2}$, respectively  the probe and the control beam detunings.
While propagation effects for the control field
are neglected the evolution of the system is
described by the following  set of Bloch equations
\begin{eqnarray}
{\rm i}\dot{\sigma}_{aa}&=&
-\Omega_{1}\sigma_{ba}+\Omega_{1}^{*}\sigma_{ab}
-\Omega_{2}\sigma_{ca}+\Omega_{2}^*\sigma_{ac}
-{\rm i}(\Gamma_{ab}-\Gamma_{ca})\sigma_{aa},\nonumber\\
{\rm
i}\dot{\sigma}_{bb}&=&-\Omega_{1}^{*}\sigma_{ab}+\Omega_{1}\sigma_{ba}
+{\rm i}\Gamma_{ab}\sigma_{aa},\nonumber\\
{\rm
i}\dot{\sigma}_{cc}&=&-\Omega_{2}^{*}\sigma_{ac}+\Omega_{2}\sigma_{ca}
-{\rm i}\Gamma_{ca}\sigma_{aa},\nonumber\\
{\rm i}\dot{\sigma}_{ab}&=&(\delta_{1}-i\gamma_{ab})\sigma_{ab}
-\Omega_{1}(\sigma_{bb}-\sigma_{aa})-\Omega_{2}\sigma_{cb},\\
{\rm i}\dot{\sigma}_{bc}&=&(\delta_{2}-\delta_{1}-{\rm
i}\gamma_{bc})\sigma_{bc}+\Omega_{2}
\sigma_{ba}-\Omega_{1}^{*}\sigma_{ac},\nonumber\\
{\rm i}\dot{\sigma}_{ac}&=&(\delta_{2}-{\rm
i}\gamma_{ac})\sigma_{ca}
-\Omega_{2}(\sigma_{cc}-\sigma_{aa})-\Omega_1\sigma_{bc}.\nonumber
\end{eqnarray}
Parameters $\Gamma_{ij}$, $i\neq j$,  describe damping of exciton states and
 are determined by exciton-damping mechanisms comprising,
temperature-dependent homogeneous broadening due to phonons and broadening
due to structural imperfections and eventual impurities.
The relaxation damping rates for the
coherence are denoted by  $\gamma_{ij}\approx\Gamma_{ij}/2$, $i\neq j$ \cite{Artoni}.
It should be noted that in the above equations only the relaxations
inside the three level system are considered,
so the total probability for the populations of the three levels is
conserved: $\sigma_{aa}+\sigma_{bb}+\sigma_{cc}=1$.
In order to study propagation of the  signal field inside the medium
the  Bloch equations are accompanied by the
Maxwell propagation equation for the Rabi frequency $\Omega_1$
of the probe pulse, which reads in the slowly varying
envelope approximation
\begin{equation}\label{kappa1}
\left(\frac{\partial}{\partial t}+c\frac{\partial}{\partial
z}\right)\Omega_1=-\mathrm{i}\kappa_1^2\sigma_{ba},
\end{equation}
where $\kappa_1^2=\frac{N\vert d_{ab}\vert ^2\omega_1}{2\hbar\epsilon_0},$
$N$ being the density of excitons, $d_{ab}$ is the transition dipole
matrix element, $\omega_1$ is the electromagnetic wave frequency.

In the first order perturbation with respect to the probe field,
the evolution of our system reduces
to the set of the following equations for the density matrix

\begin{eqnarray}
\label{blochL} {\rm i}\dot{\sigma}_{ab}&=&(\delta_1-{\rm
i}\gamma_{ab})\sigma_{ab}-\Omega_1-\Omega_2\sigma_{cb},\\
{\rm i}\dot{\sigma}_{bc}&=&(\delta_2-\delta_{1}-{\rm
i}\gamma_{bc})\sigma_{bc}+\Omega_2\sigma_{ba}.
\end{eqnarray}
Taking into account that for a weak probe pulse
polarization of the medium  for a given frequency is proportional to
the signal field $\varepsilon$ and to the susceptibility $\chi$ it has a form
\begin{equation}
 P_{1}=\epsilon_0\chi(\omega)\varepsilon_1(\omega)=Nd_{ba}\sigma_{ab}.
\end{equation}
The steady state complex exciton susceptibility exhibited to the probe field
 has the form
\begin{equation}\label{chiladder}
\chi(\omega)=\frac{N\vert
d_{ba}\vert^{2}}{\hbar\epsilon_0\Omega_{1}}\sigma_{ab}=
-\frac{N\vert
d_{ab}\vert^2}{\hbar\epsilon_0}\frac{1}{{\omega-\delta_{1}+{\rm
i}\gamma_{ab}}
-\frac{\vert\Omega_2\vert^2}{\omega-\delta_{1}+\delta_{2}+{\rm
i}\gamma_{bc}}},
\end{equation}
where  $\epsilon_0$ is the vacuum dielectric constant. The
susceptibility is a complex, rapidly varying function of $\omega$,
its real part describe the dispersion and imaginary show the
absorption
 of the medium.  Due to the dependence  of
 the refractive index $n(\omega)=\sqrt{1+\mathrm{Re}\,\chi(\omega)}$ on electric susceptibility
 we define the group index
\begin{equation}\label{group_index}
n_g(\omega)=c/v_g=1+\frac{\omega_1}{2}\frac{\partial}{\partial\omega}\mathrm{Re}\,\chi(\omega),
\end{equation}
which account on time delay of a  pulse  propagating in a medium with the group velocity
$$v_g=\frac{c}{1+\frac{1}{2}\mathrm{Re}\,\chi(0)+\frac{\omega}{2}\frac{\partial \mathrm{Re}\,\chi(0)}{\partial\omega}},$$
so the slope of dispersion inside the transparency window determines the propagation
of the pulse inside the medium. Note that derivative of the real part of the susceptibility may be positive (normal dispersion) or negative (anomalous dispersion); in the latter case the group velocity may  even become negative.\\
Another way of explaining electromagnetically induced transparency applies the notion of dressed states. Consider the
subspace spanned by the states a and c (the energy $E_c$ of the latter being moved by the photon energy $\hbar \omega_2$), coupled by
the interaction $\Omega_2$. The dressed states are eigenvectors of the Hamiltonian restricted to this subspace. The eigenenergies are
shifted from their bare values; if the control field is at resonance the shift is equal to $\pm\Omega_2$. If the probe photon is tuned right
in the middle between the dressed eigenenergies, the transition amplitudes from the state b interfere destructively.\\

In the case of spectrally not too wide pulses one can approximate the susceptibility to the lowest term of its Taylor expansion at the line center and in such a case the probe pulse can has the form
\begin{equation}
\epsilon_1(z,t)=\exp\big(i \frac{\omega_1\chi'(0)z}{2c}-\frac{\omega_1\chi''(0)z}{2c}\big)\epsilon_1(0,t-\frac{z}{v_g}).
\end{equation}
This means that the pulse moves with the velocity $v_g$ with its shape essentially unchanged apart from an exponential modification of its height  and an overall phase shift. The group velocity is approximately the velocity of the pulse maximum (exactly if there is no damping).

\section{Numerical results}
\begin{figure}
a)\includegraphics[width=.45\linewidth]{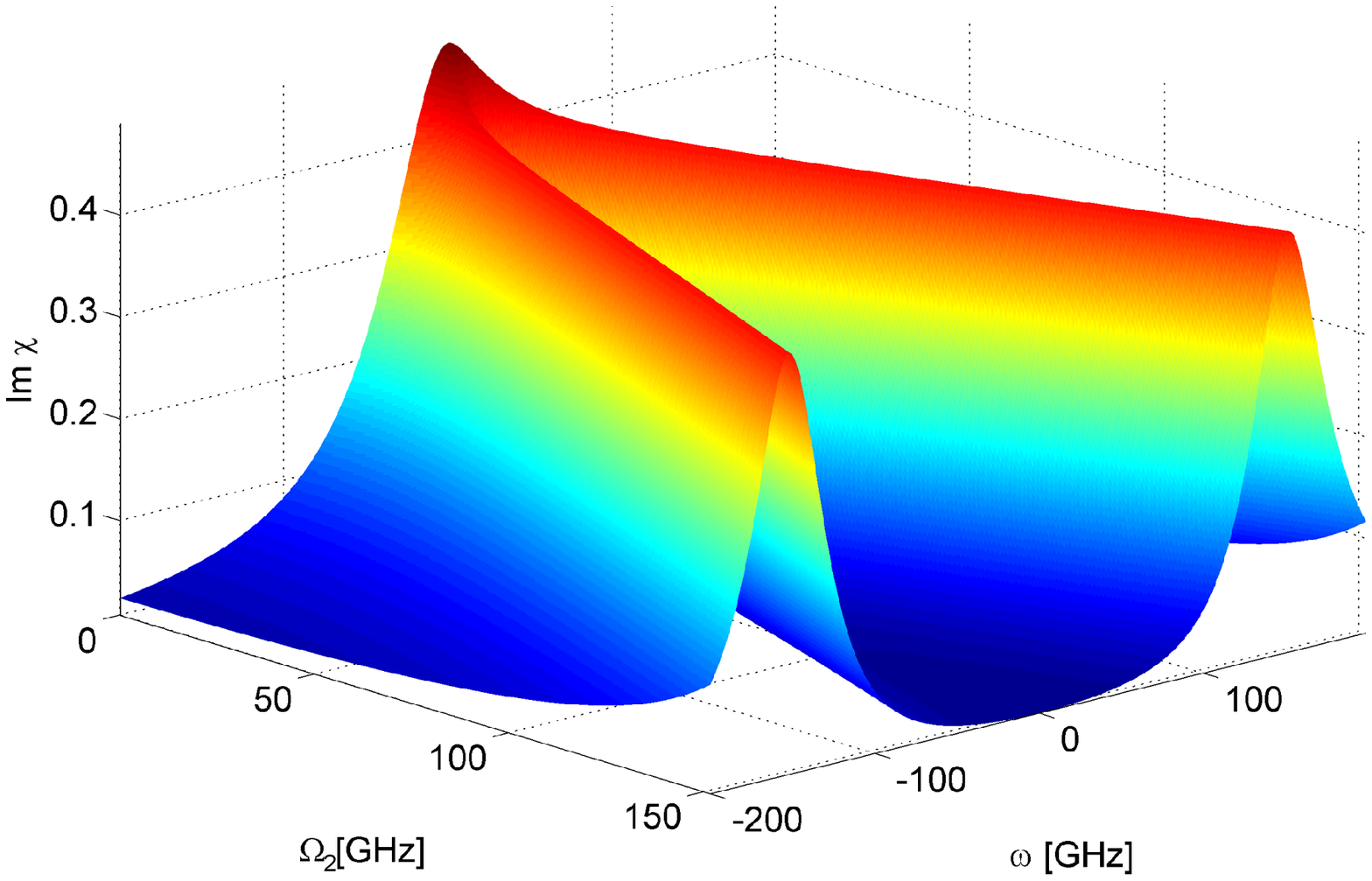}
b)\includegraphics[width=.45\linewidth]{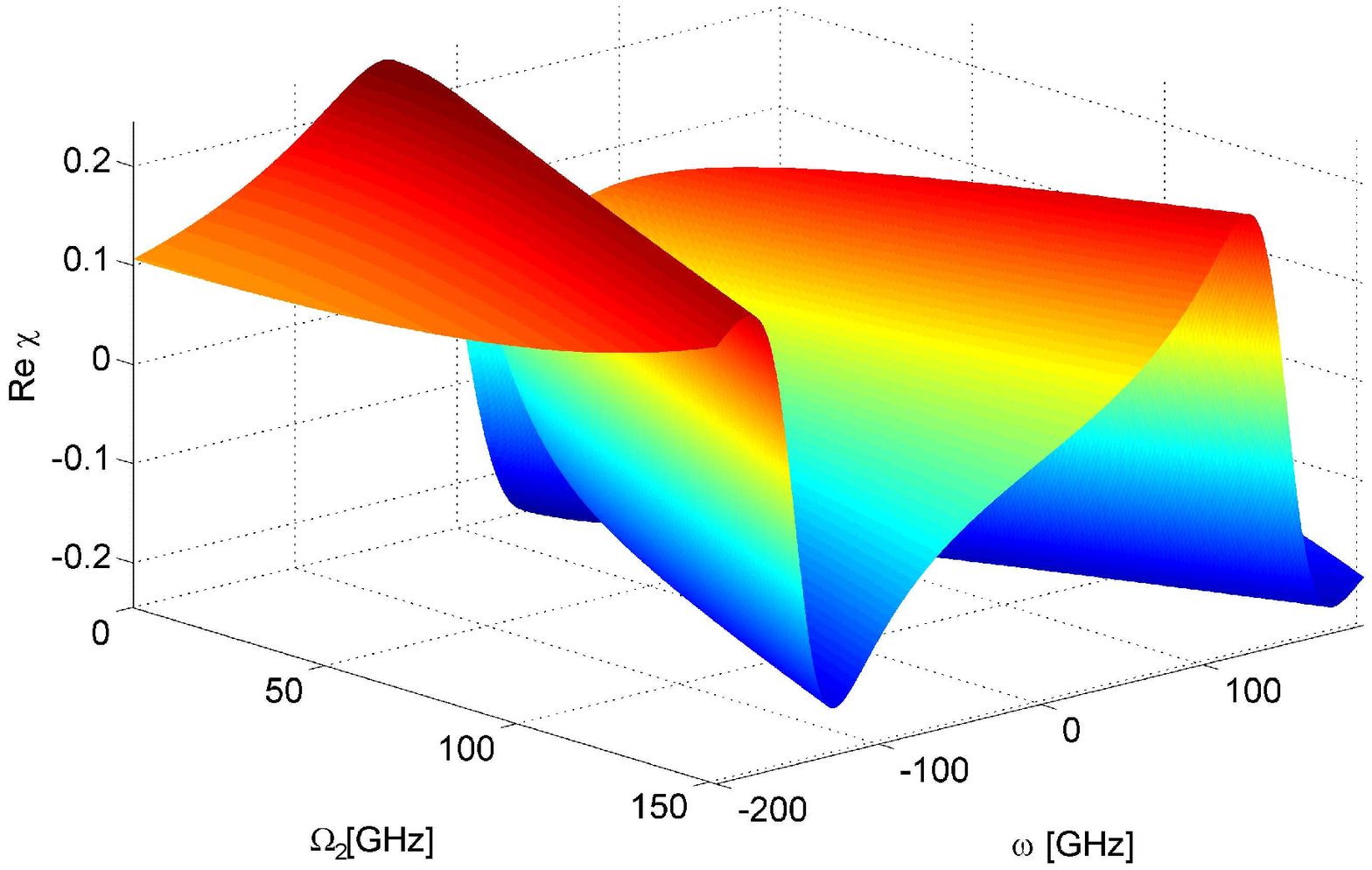}
c)\includegraphics[width=.45\linewidth]{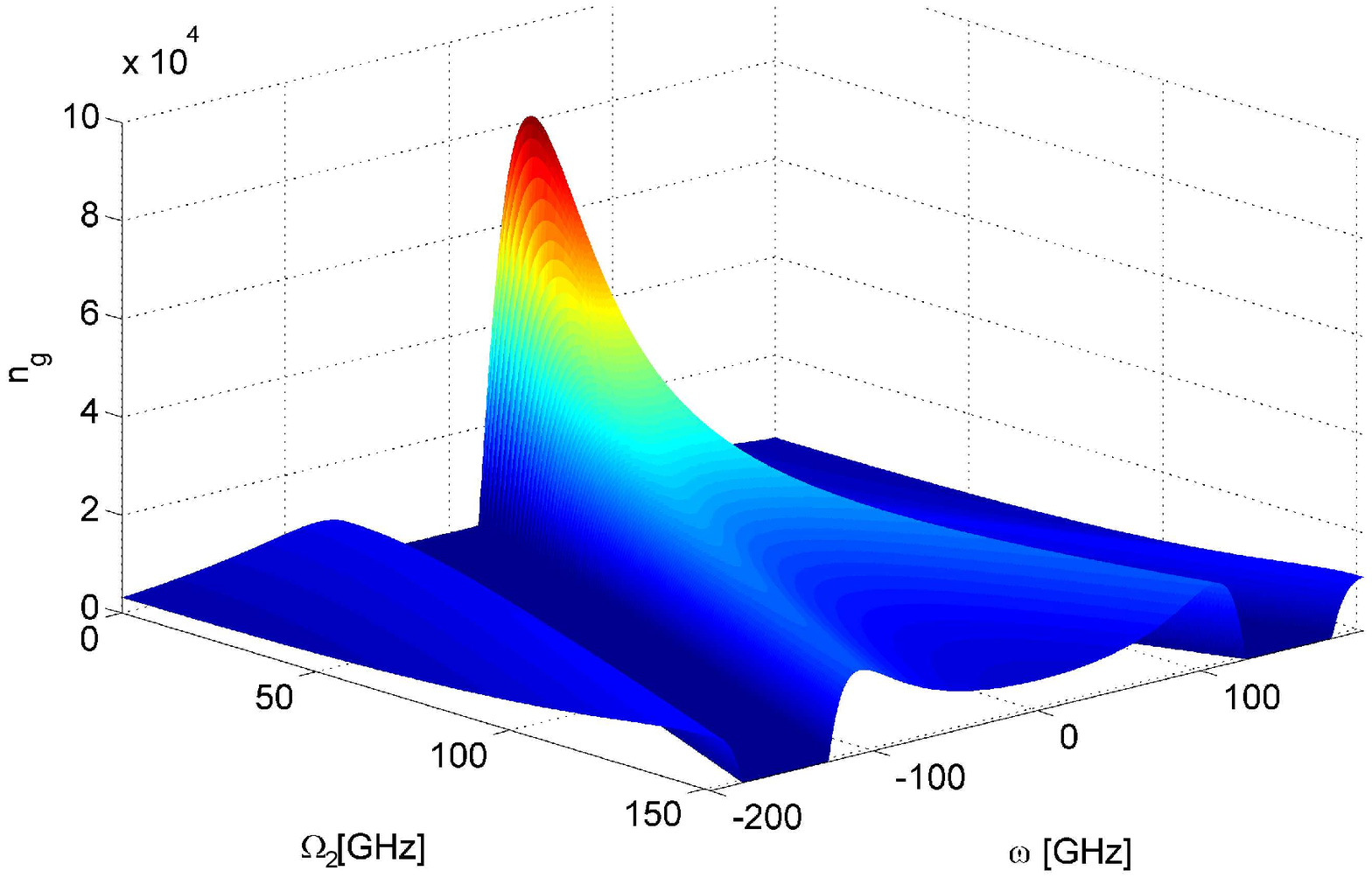}
d)\includegraphics[width=.45\linewidth]{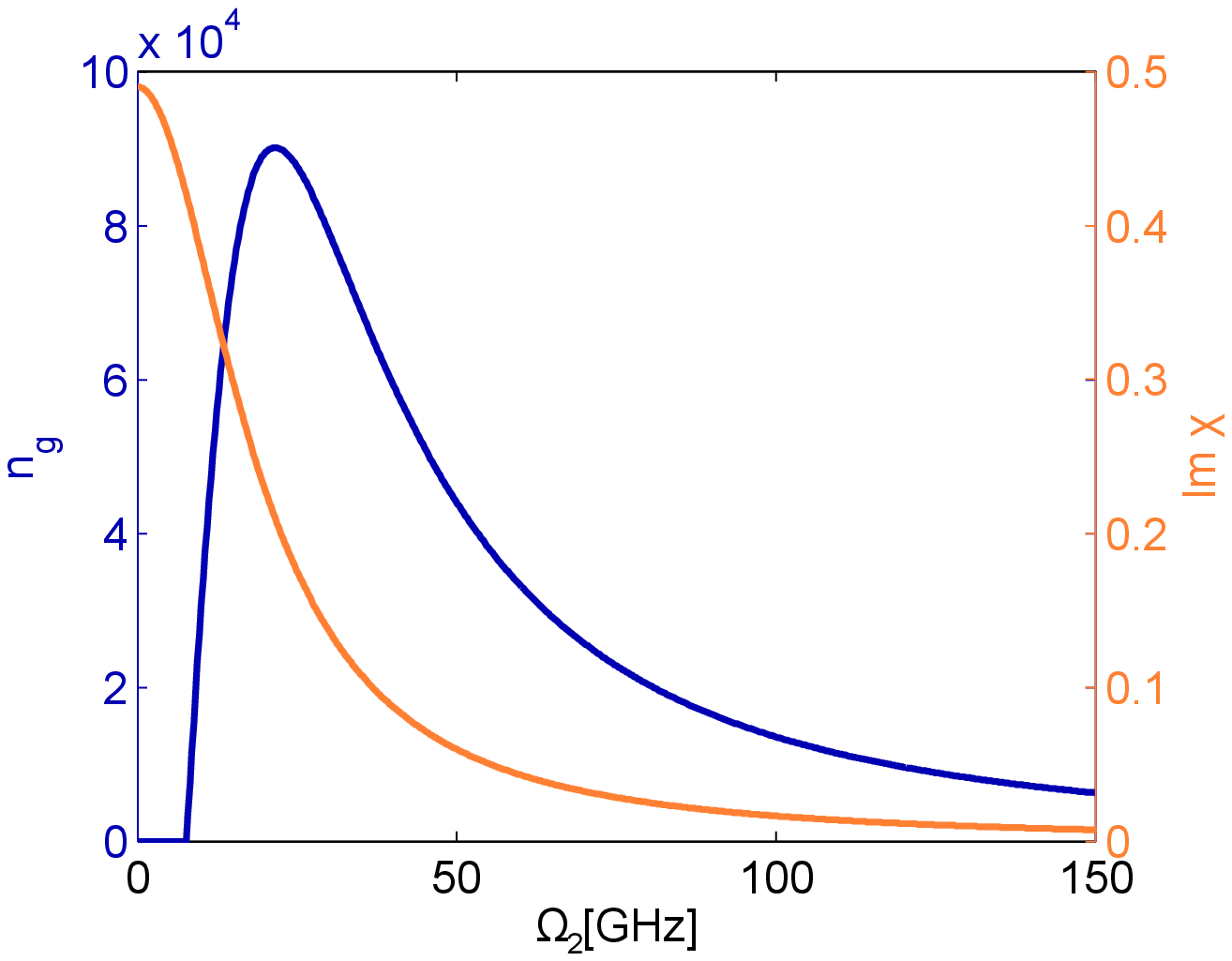}
\caption{Selected variables as a function of Rabi frequency of the control field $\Omega_2$.  a) Group velocity index $n_g(\omega)$. b) Real part of susceptibility $\chi'(\omega)$. c) Imaginary part of susceptibility $\chi"(\omega)$. d) Group index $n_g(0)$ and $\chi"(0)$, in the center of the transparency window.}\label{Fig_chi}
\end{figure}

We have performed numerical calculations for a  Cu$_2$O crystal
slab with a thickness 30 $\mu\hbox{m}$. A constant electric field
$F$ is applied in the $z$-direction.
Using the formulas
(\ref{chiladder}) and (\ref{group_index}) we have calculated the
real and imaginary part of the susceptibility and the group index.
The values of  certain energies, dipole moments
and damping parameters characteristic for the excitonic states of
Fig. \ref{Fig_schemat}  the results from our previous paper
\cite{Zielinska.PRB.2016.b} have been used. The detailed
calculations of  $N, \vert d_{ba}\vert
^2, \gamma_{ab}, \gamma_{bc} \delta_1, \delta_2$ are presented in the Appendix.
 For $N$ being the density of excitons we have used
the value $N=6.2422\cdot 10^{19}~\hbox{cm}^{-3}$.
Applying external electric field $F15~\hbox{V}/\hbox{cm}$ one obtains the following values of parameters\\
\noindent
\begin{center}
$\omega_{ab}$=3266.576~THz (=2150.3~meV)\\
$\omega_{ac}=31.402$~THz (=20.6714~meV)\\
$\gamma_{ab}=45.573$~GHz~(=30~$\mu$eV/$\hbar$)\\
$\gamma_{bc}=7.596$~GHz~(=5~$\mu$eV/$\hbar$)\\
$N=6.2422*10^{19}$ $cm^{-3}$\\
$M_{01}=0.334*10^{-60}$
\end{center}
 As can be seen in Fig.
\ref{Fig_chi} a) for non-zero Rabi frequency of the control field
the imaginary part of the system's susceptibility
 reveals a dip in the Lorentzian absorption profile  called a transparency window.
This means that a resonant probe beam which otherwise would be strongly absorbed,
is now transmitted almost without losses.
The width of the transparency window is proportional to the
square of control field amplitude and therefore increasing the control
field strength it is possible to open it out.
The real part of susceptibility is shown in Fig. \ref{Fig_chi} b).
The dispersion inside transparency window becomes normal
with the slope which increases from a decreasing control field.
The normal dispersion inside the window is responsible for
reduction of the group velocity
The absorption at the resonance does not reaches zero due to
finite value of the relaxation rate $\gamma_{cb}$, but
it is very small. This means that the medium has become
transparent for a probe pulse which travels with a reduced velocity.
It is well known that EIT allows us to obtain
a steeper slope of the refractive index and thereby a large group index, see Fig. \ref{Fig_chi} c).
The negative values of $n_g$, corresponding to the regions of anomalous dispersion and high absorption, have been omitted for clarity.
The Fig. \ref{Fig_chi} d)
shows how the group index and absorption inside the transparency window depend on the Rabi frequency $\Omega_2$.
The plots represent a cross - sections of Fig. \ref{Fig_chi} a) and Fig. \ref{Fig_chi} c) at $\omega=0$.
 For $\Omega_2=0$, the susceptibility given by Eq. (\ref{chiladder}) has a single resonance, so that the imaginary part of susceptibility is significant.
  As the Rabi frequency $\Omega_2$ increases, a transparency window is formed.
 For considered excitonic transitions optimal slowing down of order $10^{-4}$
appears for transparency window of width of tens  GHz; as shown on the Fig. \ref{Fig_chi} d),
there is some optimum $\Omega_2 \approx 25~GHz$ where the group index has a maximum value, corresponding to a narrow,
but fully formed transparency window with significant dispersion $\frac{\partial Re \chi}{\partial \omega}$. By increasing the $\Omega_2$ further, one obtains a wider window, characterized by higher group velocity but also smaller absorption.

It should be stress that the
frequency of the signal to be slowed down or even processed determines the choice
of the exciton level involved $a$. The influence of the damping rate $\gamma_{ab}$
on the group index and on the group velocity is not crucial.
Its increase causes the widening of transparency window which is accompanied
by decrease of the dispersion.

\section{Conclusions}
The universal Rydberg nature can  be exploited in systems other than atomic gases.  The impressive observation of a Rydberg blockade shift on a very different platform offers a new approach for studying semiconductor systems and also provides entirely new  long-term perspectives for developing novel devices, which are more robust and compact than atomic systems.
We have indicated the optimal states and well justified parameters to
attempt the observation EIT in Rydberg excitons Cu$_2$O media which allow to obtain considerable value of group index.
Due to coherence properties of Rydberg excitons the  manipulations of the medium transparency is possible; the width of the window and slowing down the group velocity of the pulse travelled inside the sample might be  changed in controlled way by the strength of control field.
The ability to control on-demand group index enables storing and retrieving light pulses, which is a basis to quantum memory implementation. The way of a precise dynamical control of the optical properties of the medium
by optical means  reveals new aspects of excitons quantum optics and is supposed to lead to constructing efficient tools for photonics,
e.g., delay lines, quantum switches or multiplexers.
So far experimental demonstration of EIT in Rydberg excitons media is
probably difficult to realize, but it seems safe to expect such experiments in the future.
Performing EIT in excitonic Rydberg media will be the step toward realization of
controlled interaction of Rydberg excitons in integrated and scalable solid state devices.

\appendix
\section{Energy values for S excitons}\label{Appendix A}

For the 2$P$ exciton we take the 2$Pz$ state, i.e. $\vert
2,1,0\rangle$ state, which couples with the $z$- component of the
electromagnetic wave. So all the waves propagating in the
considered medium must have the $z$- component. According to the
notation of Ref. \cite{Zielinska.PRB.2016.b} the energy of 2$Pz$
exciton follows from the
relations \\
\begin{eqnarray}\label{energy2P}
W_{200}W_{210}-\left(V_{010}^{(2)}\right)^2=0,
\end{eqnarray}
with
\begin{eqnarray}
 &&W_{n\ell m}=E_g+E_{n\ell
m}-\hbar\omega-{\rm i}{\mit\Gamma}=\nonumber\\&&=E_{Tn\ell
m}-E-{\rm i}{\mit\Gamma},\nonumber\\
\nonumber
\end{eqnarray}
where for $V^{(n)}_{010}$ and  in units $eFa^*$, one obtains
\begin{eqnarray}\label{v010}
&&V^{(n)}_{010}=\frac{1}{\sqrt{3}}\sqrt{\frac{(n-1)!(n-2)!}{16
n!(n+1)!}}\int\limits_0^\infty {\rm d}x\,e^{-x}x^4
L^1_{n-1}(x)L^3_{n-2}(x)\nonumber\\
&&=-\sqrt{\frac{12}{n^2(n^2-1)}}{n\choose n-2}{n+1\choose n-1}.
\end{eqnarray}
The energy eigenvalues have the form
\begin{equation}\label{energies1}
E_{n\ell m}=-\frac{\eta_{\ell m}^2}{n^2}R^*,
\end{equation}
where $n=1,2,\ldots, \ell=0,1,2,\ldots n-1, m=-\ell,-\ell+1,\ldots
+\ell$, with $\eta_{\ell m}$ defined by with
\begin{equation}\label{etaellm}
\eta_{\ell m}=\int\limits_0^{2\pi} {\rm d}\phi\int\limits_0^\pi
\sin\theta\,{\rm d}\theta\frac{\left|Y_{\ell
m}\right|^2}{\sqrt{\sin^2\theta +\gamma^2
\cos^2\theta}},\end{equation} $\gamma=\mu_\parallel/\mu_z$ being
the exciton effective masses anisotropy parameter, $L^m_n$ are the
Laguerre polynomials, and $Y_{\ell m}$ being the spherical
harmonics. The energy for the $2P$ exciton results from Eq.
(\ref{energy2P}) where we take the larger value from the two
solutions $E_{1,2}$.
\\The energy for $10S$ exciton could be calculated
from the equation
\begin{equation}
W_{10,00}W_{10,10}-\left(V_{010}^{(10)}\right)^2=0,
\end{equation}
where we take the smaller value from the solutions. The dipole
matrix element $\vert d_{ba}\vert ^2$ follows from the relation
\begin{eqnarray}
&&\left|M_{10}\right|^2=\frac{4\epsilon_0\epsilon_ba^{*3}\Delta_{LT}^{(P)}}{\pi
(r_0/a^*)^2\eta_{11}^5},
\end{eqnarray}
$\Delta_{LT}^{(P)}$ being the longitudinal-transversal splitting,
and $r_0$ the so-called coherence radius.
\begin{figure}
a)\includegraphics[width=.45\linewidth]{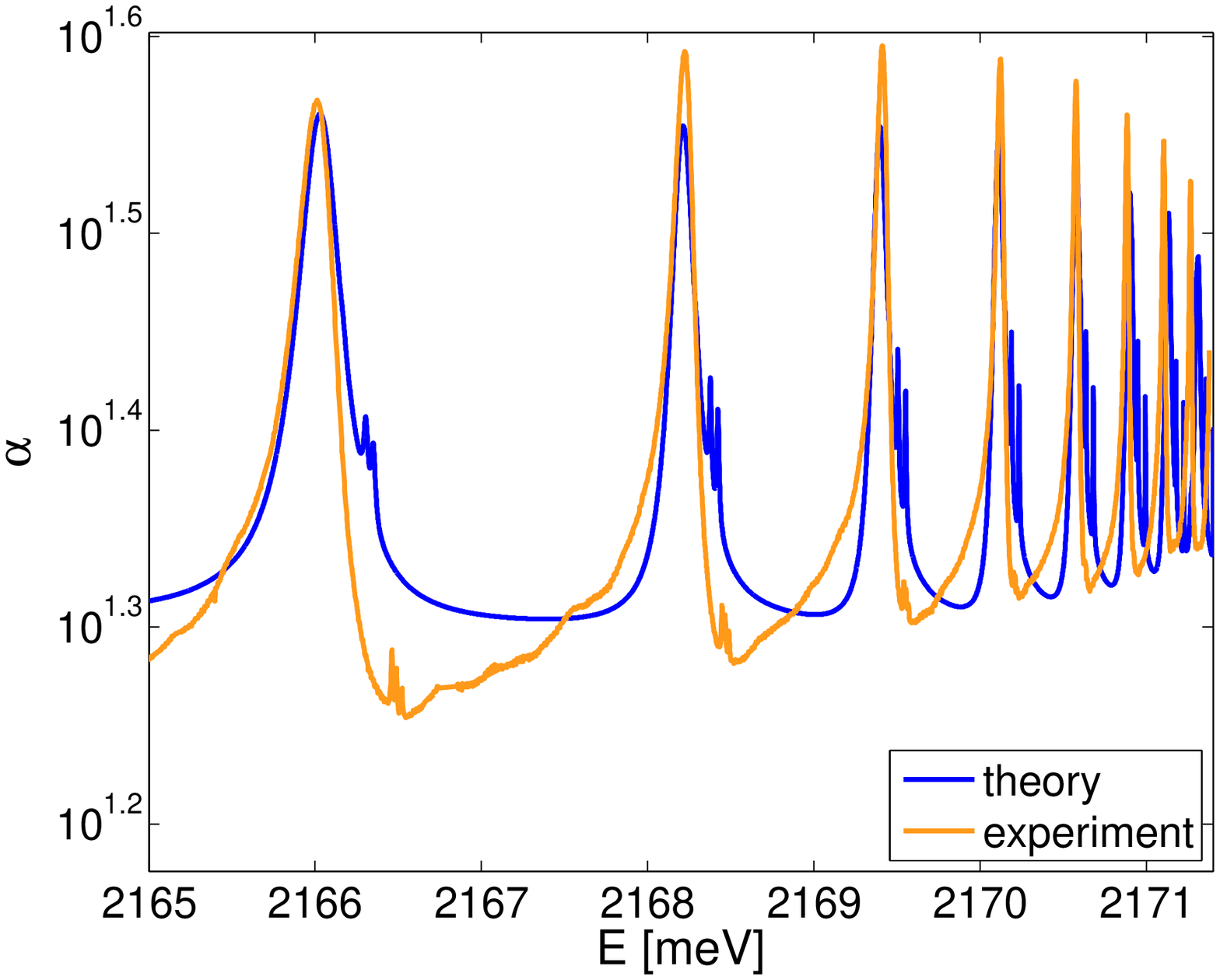}
b)\includegraphics[width=.45\linewidth]{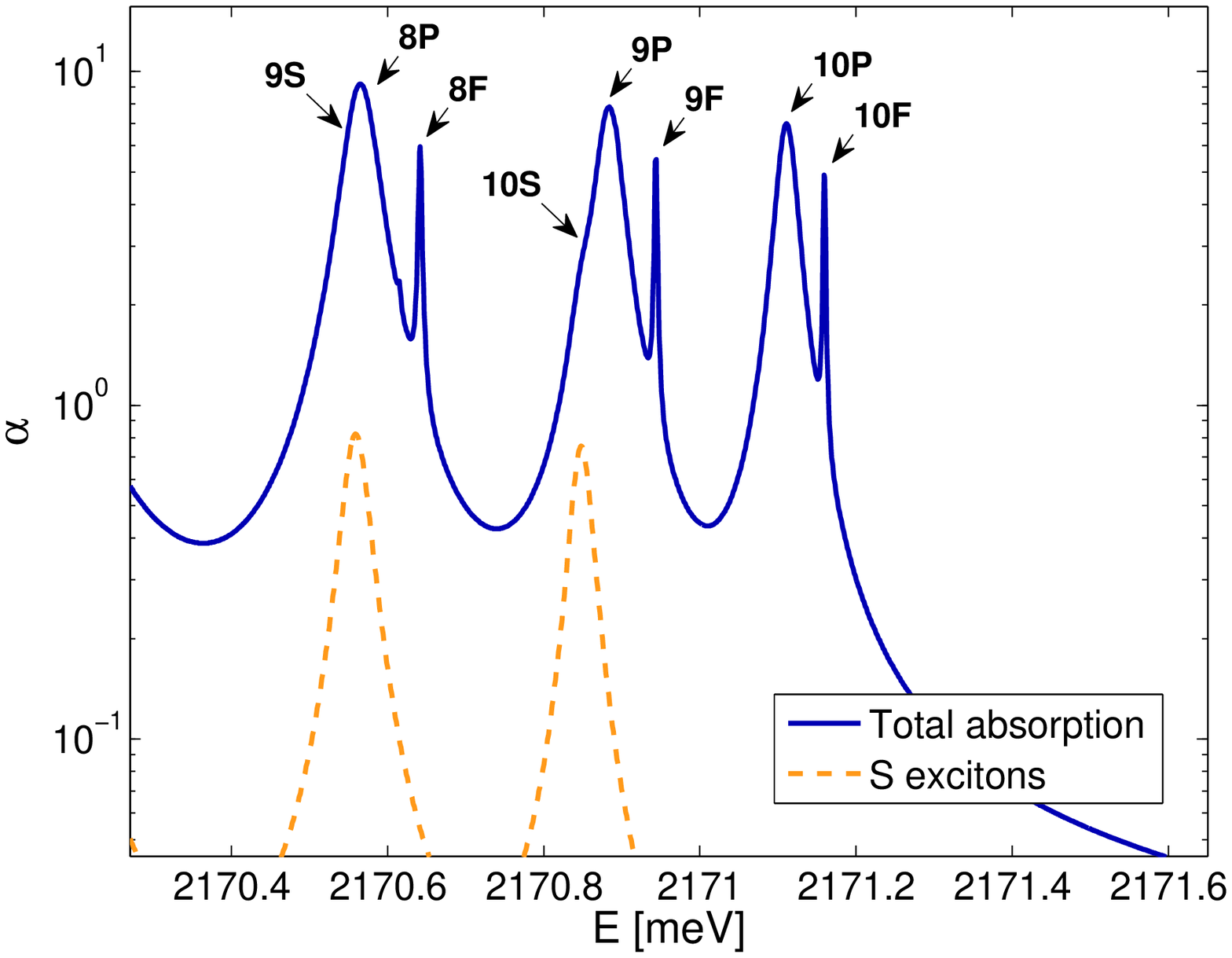}
\caption{The absorption spectrum of the considered system, for principal quantum number n = 4 - 10. a) Comparison of the obtained results with experimental data \cite{Heckotter}. The line widths provide the estimation of damping constants $\gamma$. b) Detailed spectrum near n = 10, showing the overlap of S and P excitonic lines.}\label{Fig_comp}
\end{figure}
Positions of our resonances were in perfect
agreement with the experimental data (see Fig. \ref{Fig_comp} a)
The damping parameters were determined by fitting our results
regarding to the linewidths of resonances. We were able to
estimate the dissipation rates of the P excitons (Fig.
\ref{Fig_comp} a) and S excitons (Fig. \ref{Fig_comp} b) which we
later used in our numerical calculations. The relaxation rates are
comparable to the values found in prior literature
\cite{Kazimierczuk}.

\end{document}